\documentclass[final, pra, a4paper, english, 11pt, twocolumn, notitlepage, floatfix]{revtex4-2}

\usepackage{lmodern}

\usepackage[utf8]{inputenc}
\usepackage[T1]{fontenc}

\usepackage{esvect}

\usepackage[english]{babel} 
\usepackage{textcomp}
\usepackage{setspace}
\usepackage[USenglish]{isodate}
\usepackage[per-mode=symbol,separate-uncertainty]{siunitx}
\DeclareSIUnit \belm {Bm}
\DeclareSIUnit \eurom {\euro}

\pdfoutput=1    
 
\usepackage{graphicx}
\usepackage{amsmath, amssymb, amsfonts,bm}
\usepackage{latexsym} 
\usepackage[protrusion=true, expansion=true]{microtype}
\usepackage{mathtools}
\usepackage{placeins}  
\usepackage[usenames, dvipsnames]{color}
\usepackage{layout}
\usepackage{multirow}
\usepackage{tabularx}
\usepackage{tabulary}
\usepackage{tabu}
\usepackage[table, x11names]{xcolor}
\usepackage{qcircuit}
\usepackage{stackengine}
\usepackage{cancel}
\usepackage{graphicx}
\usepackage{eurosym}
\usepackage{graphicx}
\usepackage{fancyhdr}

\definecolor{navyblue}{rgb}{0.0, 0.0, 0.5}
\usepackage[colorlinks=true, breaklinks=false, linkcolor=navyblue, urlcolor=navyblue, citecolor=navyblue]{hyperref}

\DeclareGraphicsExtensions{.jpg}
\begin{document}
\title{Towards reliable quantum software, algorithm and use-case development: Multidisciplinary analysis from the perspective of Finnish industries}
\author{Matti Silveri$^1$}
\author{Tommi Mikkonen$^2$}
\author{Kimmo Halunen$^3$}
\author{Teiko Heinosaari$^2$}
\author{Aravind Plathanam Babu$^1$}
\author{Vlad Stirbu$^2$}
\author{Oskari Kerppo$^2$}
\author{Majid Haghparast$^2$}
\author{Andrés Muñoz-Moller$^2$}

\affiliation{$^1$ Nano and Molecular Systems Research Unit, University~of~Oulu, Finland\\$^2$ Faculty of Information Technology, University~of~Jyväskylä, Finland\\$^3$ Biomimetics and Intelligent Systems Group, University~of~Oulu, Finland}

\date{\today}

\begin{abstract}
Quantum computing is a disruptive technology with the potential to transform various fields. It has predicted abilities to solve complex  computational problems beyond the reach of classical computers. However, developing quantum software faces significant challenges. Quantum hardware is yet limited in size and unstable with errors and noise. A shortage of skilled developers and a lack of standardization delay adoption. Quantum hardware is in the process of maturing and is constantly changing its characteristics rendering algorithm design increasingly complex, requiring innovative solutions. Project "Towards reliable quantum software development: Approaches and use-cases" TORQS has studied the dilemma of reliable software development and potential for quantum computing for Finnish industries from multidisciplinary points of views. Here we condense the main observations and results of the project into an essay roadmap and timeline for investing in quantum software, algorithms, hardware, and business.
\end{abstract}

\maketitle

\section*{Summary}
\subsubsection*{Key recommendations for Finnish industry}
\begin{itemize}
    \item Start building quantum-classical algorithm capabilities now. Focus initially in optimization, material, and molecule simulations.
    \item Engage in practical benchmarking  of industry-specific use-cases to understand better when quantum can outperforms classical. Utilize online quantum computers, e.g., the Finnish 50 qubit computer, in practical tests. 
    \item To bridge the gap between theoretical possibilities and practical implementations, prioritize research efforts that focus on adapting and converting classical algorithms into their quantum and quantum-classical counterparts
    \item Focus near-term R\&D on high-value software stack components: middleware, high-level programming, use-case specific method development, and HPC-integration
    \item Train technical staff to critically track quantum hardware and software evolution
\end{itemize}
\subsubsection*{Highlights}
\begin{itemize}
    \item Expected quantum hardware, algorithm and software progress in 2025-2035 synthesized on roadmaps by quantum technology companies, Fig.~\ref{fig:hw}
    \item Feasibility ranking of quantum algorithms and expected early practical use-cases, Fig.~\ref{fig:alg}
    \item Identification of key development areas in quantum software, Sec.~\ref{sw}
    \item The best performing quantum software needs to be linked and optimized on the specific hardware. This is a strong contrast to contemporary classical software engineering.  
    \item  Practical quantum programming and testing gives important insights to tools and techniques. Suggestions on focusing on the problem mapping and partitioning, algorithm selection based on hardware restrictions, and classical parts of the hybrid computation.
\end{itemize}
\subsubsection*{Strategic risks of inaction and benefits of early adoption}
\begin{itemize}
    \item Hardware and software develop rapidly. Early experimentation and scaling analysis works excellently for tracking progress and increasing preparedness for timely investments
    \item Industry-academia-startup collaborations are beneficial in accessing diverse quantum expertise and de-risking early exploration.
\end{itemize}

\section{Introduction}

\begin{figure*}
    \includegraphics[width=1\linewidth]{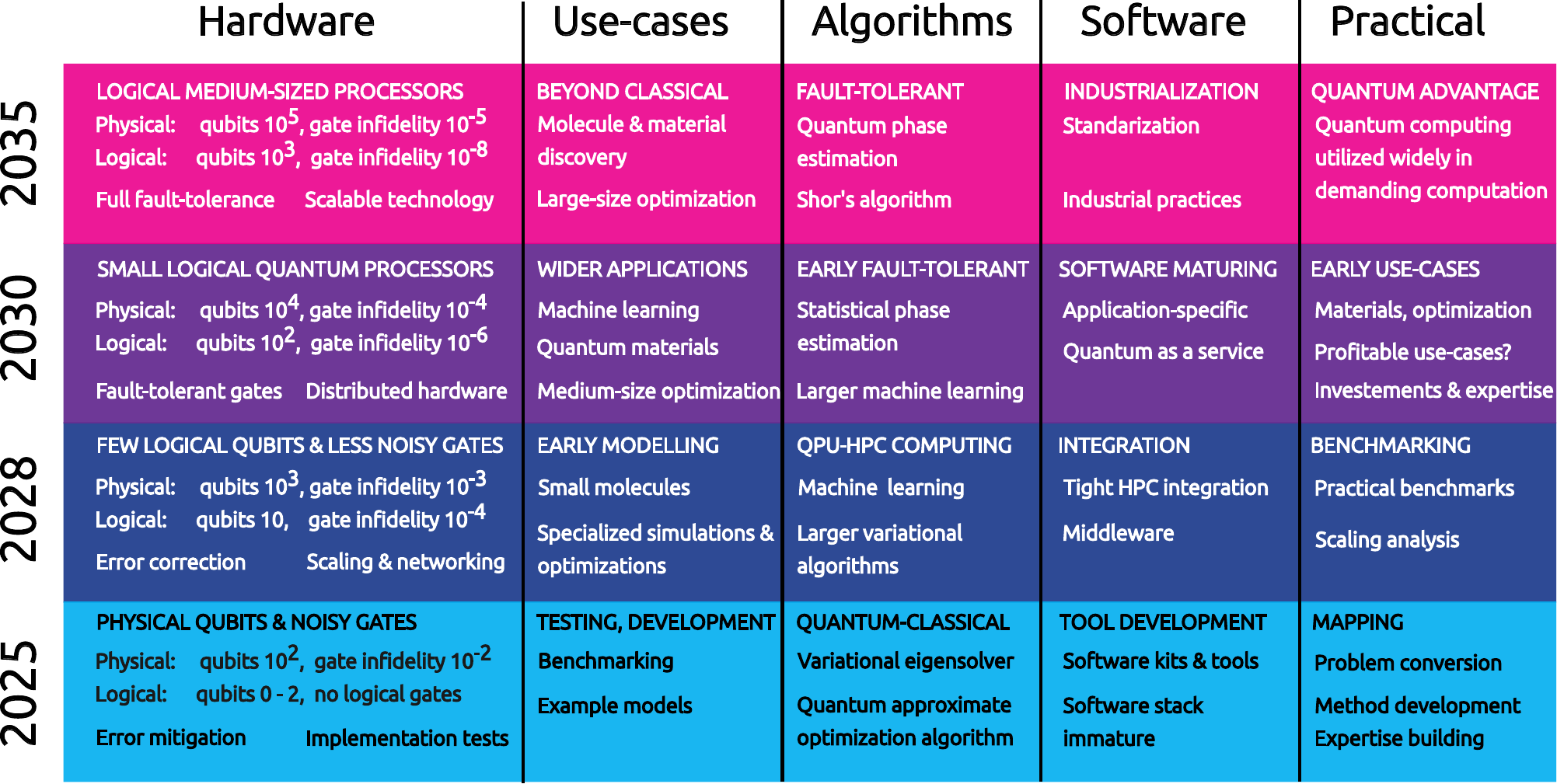}
    \caption{\label{fig:hw}Expected progression in quantum hardware, use-cases, algorithms, software and practical in 2025-2035. Table synthesizes roadmaps by various quantum technology companies~\cite{IQM25, IBM25, Google25, Pascal25, Quera25}. Gate infidelity refers to two-qubit gates.}
\end{figure*}

It has been widely recognized that quantum computing has major potential for disruptive effects on many fields. Finnish enterprises are aware of this potential, but translating the awareness into action requires answering the following major question: 
\begin{quote}
    \emph{When and to which aspects of quantum computing to invest?}
\end{quote}
Quantum computing mixes several disciplines: quantum physics, computer science, software engineering, applied mathematics and various specific use-case areas. In this light, we examine the main question through the subquestions: 
\begin{itemize}
    \item[(a)] \emph{When will quantum hardware be mature enough for company-specific use-cases?}
    \item[(b)] \emph{When is the right time to invest in quantum software development?}
    \item[(c)] \emph{Which parts of the quantum computing stack will provide the most added value now and near-future?}
\end{itemize}
This roadmap and timeline is a synthesis of results of the research and development project “Towards reliable quantum software development: Approaches and Use-cases (TORQS)” in 2023-2025. Funded by Business Finland, the research was carried out at the University of Oulu and the University of Jyväskylä in collaboration with nine Finnish companies Algorithmiq, Futurice, Loihde Advance, 61N Solutions, OP Financial Group, Orion Pharma, Quanscient, QuantrolOx and Solita.

Several Finnish and international stakeholders have prepared or will publish quantum technology roadmaps, strategy papers or agendas~\cite{FinnishQuantumAgenda, QuantumStrategy_Finland25}. Our report complements them by offering concise research-based observations and suggestions. The target audience is Finnish industry and decision makers in education, research and development.

We first review four main pillars of quantum computing in Secs.~\ref{hw}-\ref{sw}: hardware, algorithms, use-cases and software. Then we discuss education of quantum experts in Sec.~\ref{edu}. We conclude by elaborating the Finnish industry and enterprise perspective to quantum computing from mid-2020's to mid-2030's in Sec.~\ref{ind}.

\section{Quantum hardware}\label{hw}
The quantum hardware industry is in a rapid development stage. Focus shifts from increasing qubit counts to achieving high-quality qubits with improved coherence times and lower error rates in quantum gates. Hardware roadmaps highlight the importance of pushing gate-error levels down to achieve error-corrected quantum hardware and its practical integration with classical supercomputers to develop hybrid computing architectures~\cite{IQM25, IBM25, Google25, Pascal25, Quera25}. Scalability requires functional modularity and distributed quantum architectures through interconnects and networks, see Fig.~\ref{fig:hw}. 

Finnish expertise and activities are mainly currently in superconducting circuit based quantum technologies. Globally looking several physical platforms are in fierce competition. Superconducting circuits benefit from fast gates and scalable fabrication. Neutral atom quantum computers have presented impressive qubit numbers together with first quantum error correction protocols. Strong investments have been made in photonic approaches, but major scientific demonstrations are yet to come. Trapped ion platforms have one of the best gate errors, and have already put forward impressive error correction capabilities. Spin qubits have recently demonstrated improved gate accuracies. 

The quantum computers in mid-2020's are clearly not yet useful for  practical use-cases. This is visible in the roadmaps of major hardware manufacturers. One possible future approach is to combine technologies for reaching complementing hybrid quantum technologies. Also it can be possible that different technologies mature at very different pace.  We are still in the era of building foundation blocks. 

As quantum software and hardware are closely linked, realizing hardware-agnostic software is very challenging. The best performing quantum software needs to be linked and optimized on the specific hardware, and sometimes even for the specific hardware versions. This is a strong contrast to contemporary classical software engineering.  

\section{Theoretical algorithm and use-case analysis}\label{alg}
Quantum computing can already be studied in practical terms in some use cases as an aid for solving relevant problems. Here the word 'use-case' means a specific application area or a problem. It does not mean that there is an advantage for utilizing quantum computing methods. Actually, as of 2025, there has not yet been practical demonstrations of quantum advantage in practical use-cases. 

\begin{figure}[]
    \centering
    \includegraphics[width=1.0\linewidth]{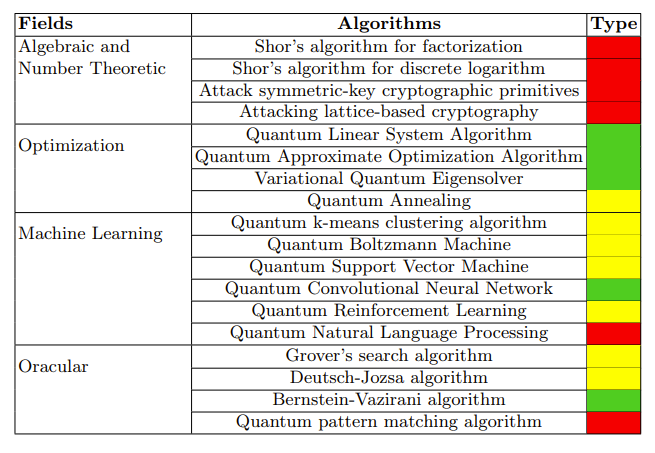}
    \caption{\label{fig:alg}Feasibility ranking of quantum algorithms and use-case areas~\cite{Bui24}:  Green means expected near-term feasibility, yellow possible medium-term applicability, and red that the algorithm will clearly need over a decade of advances in order to be applicable. }
\end{figure}

Based on our research we have identified some use-cases areas where advantage of quantum computing is closer to practical than with others. The analysis is based on a wide literature review and analysis on the structure of the quantum algorithms~\cite{Bui24}. We have ranked the use-case and algorithm classes by traffic light colors: green means demonstrated and expected near-term feasibility, yellow possible medium-term applicability, and red that the algorithm will clearly need  over a decade of further advances in order to be applicable.

Fig.~\ref{fig:alg} shows the ranking of some relevant quantum algorithms. At the moment it seems that optimization and simulation algorithms are most likely to be feasible in the near term. Also machine learning algorithms show promise for the longer term. The algorithms threatening the security of current cryptographic solutions still luckily seem to be quite far in the future.

One key aspect of understanding the potential of quantum algorithms is that in many use cases these quantum algorithms are only a part of a more complex system that is used to achieve some higher level goal e.g. a portfolio optimization. Thus, the most probable use cases for quantum algorithms will be in the hybrid mode, where both classical and quantum computing hardware, software and algorithms are used. Analyzing the potential of this approach needs to be taken into account when evaluating the full potential of quantum computing.

Beyond just algorithms, we should also explore quantum data: development and utilization of quantum random access memories and efficient reading and writing of quantum registers. Also quantum simulation e.g. for material science is a potential use-case area. This could mean simulating dynamics of a complex quantum system on a quantum computer, for example molecules relevant to the pharmaceutical industry, and then immediately using the obtained data to learn about the quantum system itself. 

\section{Practical use-case analysis}\label{use-cases}
The project was shaped with various use cases where practical implementations have been composed. In more detail we studied optimization problems due to their wide industrial applicability: process optimization, logistics, finance etc. Practical programming and testing gave important insights to tools, techniques, and industry expectations~\cite{Liimatta24}.

While current noisy intermediate-size quantum computers have not yet achieved fault-tolerance or the scale necessary for quantum advantage, they serve as a valuable platform for experimentation.  Dominant designs have started to emerge in terms of quantum ecosystems as well as technological concepts. They enable the advancement of next-generation hardware, the development of quantum algorithms, and the validation of quantum technologies in practical applications. However they are not yet powerful enough to run industry-scale solutions. 

Our practical experiments suggest focusing in three main phases in implementation of use-cases in quantum computers: problem mapping and partitioning, quantum algorithm selection considering the quantum processor limitations and classical parts of the hybrid quantum-classical algorithms. This also gives an suggestion how a use-case team should be composed: There should be persons such that all these expertise areas are solidly covered. 

\emph{Problem mapping and partitioning:} The use-case needs to be mapped to a form that can be then solved with a quantum algorithm. This mapping consists of two parts. First, it requires the description of the the equations and mathematical model. Second, these equations have to be transformed into format that can be modeled by using qubits. Since this typically leads to large systems with excessively many qubits, an important factor is partitioning and approximating of the problem. This approach involves classical-quantum partitioning, where only the most computationally demanding or quantum-suitable parts of the problem are delegated to the quantum computer, while the rest are handled by classical methods.

\emph{Quantum algorithm selection based on hardware restrictions}:  When selecting a quantum algorithm, the decision is heavily influenced by the choice of hardware. Given that the algorithm depends so heavily on the hardware architecture, we suggest driving efforts into scenarios where the quantum device solves the problem of interest as naturally as possible. In practice this could mean selecting mapping and algorithm so that physical qubit topology of the processor and the resulting qubit connections of the problem match as close as possible. 

\emph{Classical parts of the hybrid algorithms}: Currently, the most suitable and feasible algorithms are hybrids between quantum and classical computations. In some sense they are analogous to the computational models of machine learning and artificial intelligence, just that the neural network is been replaced by a parameterized quantum gate sequence and whose parameters are optimized, that is, trained via classical optimizers. To build efficient software, one needs to master and optimize both quantum and classical algorithms here.

\begin{figure*}
    \centering
    \includegraphics[width=0.66\linewidth]{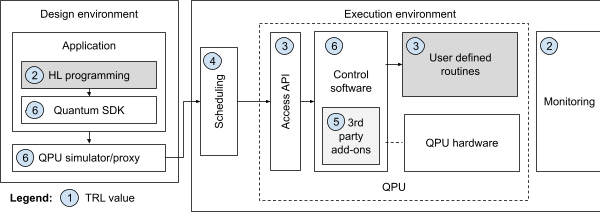}
    \caption{\label{fig:sw}Quantum software stack depicted through the major components and their current TRL levels.}
\end{figure*}

\section{Quantum software}\label{sw}
The quantum software stack is a loose concept seen as a distributed system composed of a multitude of classical and quantum software components and subsystems. Today the major efforts are around integrating quantum computing with classical high-performance computing environments. Quantum computing could similarly provide computational advantage for a multitude of applications implemented using cloud technologies. Figure~\ref{fig:sw} depicts the major components of the quantum software stack and provides their maturity level according to the technology readiness level (TRL)~scale. 

The software stack can be divided into two environments: design – controlled by the user or application developer, and execution – controlled by the quantum processor unit~(QPU) manufacturer or operator. The user designs applications that incorporate quantum components using a quantum software development kits~(SDK). Popular examples are Qiskit and Pennylane. Then user executes it on a local or remote quantum processor or on a simulator. Current quantum software development kits provide a low-level gate-based metaphor to designing programs. This poses limitations in scalability in terms of qubit and gate numbers, and expressiveness meaning ability to capture the intent of the algorithm developer. Efforts for addressing these limitations by developing high-level~(HL) programming languages have been initiated. 

The execution environment contains the software that is used to effectively run the quantum hardware, and the software infrastructure needed for scheduling. Scheduling means finding the appropriate hardware resources for the quantum routine. Monitoring means collecting data about quantum routine execution.  While the scheduling is typically achieved using high-performance or cloud computing infrastructure, developing efficient monitoring of routine execution is mostly in its infancy. 

The quantum processor implementation is realized by classical systems that convert the quantum routine, received as a quantum gate sequence a.k.a. a circuit, to the appropriate quantum hardware technology. The hardware manufacturers expose proprietary interfaces, but it is generally understood that this interface should be standardized. The circuit is converted by the control software into a sequence of hardware specific pulses. This process involves critical hardware-specific transpilation steps, such as routing and qubit mapping, to adapt the circuit to the physical topology of the quantum hardware. Third-party tools can support the quantum programming workflow by providing additional functionalities or plugins—for example, to enable  error mitigation, or enhanced transpilation strategies. Some quantum hardware manufacturers allow application developers to provide user defined routines for improved quantum error correction and mitigation.

The areas with low TRL are ripe for innovation in the following areas: application development – unified experience for high-level programming and user defined routines; specialized compilation extensions –- quantum error correction and mitigation, middleware routines and other high value steps via 3rd party plugins; effective operation – scheduling and monitoring infrastructure. Development should be targeted in open fora to enable easy integration into the larger quantum software ecosystem, pushing the big companies to maintain a competitive environment that fosters innovation.

\section{Education of quantum experts}\label{edu}
Quantum computing has traditionally been an advanced topic in physics curricula with solid foundation in quantum mechanics and mathematics. While this approach ensures a deep understanding and is essential, e.g., for algorithm and hardware development.  Quantum computing is in the process of expanding from natural sciences towards practical topics, various use-case areas, software and computer engineering. Restricting only to physics courses can act as a barrier for learners from other disciplines or those with limited prior knowledge.

First, recognizing the need for broader accessibility, the TORQS project developed a novel approach to quantum computing education by designing and implementing a quantum circuit simulator. This simulator functions as a plugin component within a learning environment (TIM in this case). Quantum circuit simulator integrated into a online learning environment opens new avenues for interactive learning. It allows educators to create exercises and activities tailored to learners of varying levels in mathematics, physics and programming. This is essential for making it accessible to a wider audience, including participants from fields like computer science, engineering, or even humanities. 

Online learning environment method emphasizes conceptual understanding and practical, interactive experimentation, enabling learners to explore quantum circuits and algorithms interactively.  For instance, students can interactively construct quantum circuits, observe their behavior, and analyze the outcomes directly within the learning environment.

In addition to broadly accessible online course and deep quantum physics courses, the project has recognized an urgent need for practical programming-oriented quantum computing courses. After completing these courses students would master popular quantum software development environments, knows how to use quantum computers and simulators, be fluent in hybrid and early fault-tolerant quantum algorithms and know how to solve simple quantum use-cases. 

\section{Industry and enterprise perspective}\label{ind}
The industrial utilization of quantum computing remains yet in its infancy despite significant progress in hardware development. While companies and research organizations have achieved milestones in creating and improving quantum processors, the translation of these advancements into practical industry applications continues to face challenges.

One of the most promising domains for quantum computing lies in optimization problems. Algorithms such as quantum approximate optimization algorithm~(QAOA) for quadratic unconstrained binary optimization~(QUBO), quantum amplitude estimation~(QAE), and variational quantum eigensolvers~(VQE) are well-suited to tackling complex optimization tasks on noisy intermediate-scale quantum hardware~\cite{dalzell23}. When scaled to large enough problems, these algorithms hold potential for industries like logistics, finance, and pharmaceuticals, where optimization-type tasks play a critical role. However, realizing these potentials demands significant advancements in quantum algorithm and hardware development. It is important to keep in mind fault-tolerant and early-fault tolerant counterparts of these algorithms, e.g., quantum phase estimation and statistical phase estimation~\cite{SPE22}, which are by construction more powerful and complicated.

To bridge the gap between theoretical possibilities and practical implementations, organizations must prioritize research efforts that focus on adapting and converting classical algorithms into their quantum counterparts. This requires not only a deep understanding of classical computational challenges but also an intimate knowledge of quantum mechanics and the unique constraints of current and future quantum hardware. Such efforts can significantly accelerate the development of industry-relevant quantum applications. Our suggestions is that industry would start preparing by acquiring technical expertise in quantum and high-performance classical computing. With these expertise, they can better follow the rapid hardware, software and algorithm development and be prepared for further quantum computing investments in the right time.  Furthermore, we see that industry-academia and industry-startup collaborations are an excellent way to access diverse quantum computing expertise and de-risk early exploration.

Another major hurdle is the current state of quantum software infrastructure. Unlike classical software ecosystems, which have mature pipelines for development, testing, and deployment, quantum software lacks the robustness required for transitioning from experimental algorithms to production-ready solutions. Establishing reliable quantum software frameworks and tool chains is crucial to overcoming this limitation. Doing so will enable industries to integrate quantum computing into their operations more effectively and at scale.

\section{Summary}\label{conc}
The roadmap and timeline is schematized in Fig.~\ref{fig:hw} gathering observation and expectations on the development in quantum computing during the next 10 years until 2035.

 \emph{When will quantum hardware be mature enough for company-specific use-cases?} One cannot give definite answer since it depends on the progress of hardware, the specific nature of the use-case, the demands of the corresponding quantum algorithm, and the performance of the current classical algorithm. In all these areas, there are diverse development tracks and technical aspects. The best approach is to do a benchmark studies to estimate the gap between the performance between the best classical and quantum solvers, and make interpolations on closing the gap. 

  \emph{Which parts of the quantum computing stack will provide the most added value now and near-future?} Different parts of the quantum software stack are in different TRL levels, ranging from 2-6. Thus now the areas with low TRL have high potential for most added value: high-level programming, middleware such as gate and qubit transpilation and routing, as well as, scheduling and monitoring. In near-future, software and algorithm development related to specific quantum computing use-cases, for example, related to quantum simulation of materials, hybrid classical-quantum algorithms for optimization and scientific and technical modeling of molecules and processes. 

\emph{When is the right time to invest in quantum software development} Quantum hardware industry has grown during the last decade strongly. Assuming the hardware progress keeps its pace, this will be followed by a growth in quantum software, similar to classical software. Thus, to build expertise, understand limitations, be ready for hardware improvements, and to influence the developing ecosystem, it is good to start building capabilities already now. 

\newpage
\section*{Acknowledgments}
We acknowledge the whole TORQS project team for active and fruitful discussions. The work is supported by Business Finland through the project "Towards Reliable Quantum Software Development: Approaches and Use Case" (TORQS), Grants No. 8582/31/2022 and 8436/31/2022. Matti Silveri also acknowledges H2Future project through the Research Council of Finland (Grant. No. 352788) and the University of Oulu, and the Research Council of Finland Grant No. 316619.

\bibliography{roadmap}
\end{document}